\RequirePackage{fix-cm}
\documentclass{svjour3}                     
\smartqed  
\usepackage{graphicx}
\usepackage{amsmath}
\usepackage[ruled,vlined]{algorithm2e}

\begin{document}

\title{Computer-Aided Generation of N-shift RWS}
\subtitle{A Computational Approach to Generate Multi-Shift Rotational Workforce Schedules}

\author{Benjamin Bolling}


\institute{B. Bolling \at
              \email{benjaminbolling@icloud.com \at
              }           
}

\date{Received: date / Accepted: date}

\maketitle

\begin{abstract}
Generating schedules for shift workers is essential for many employers, whether the employer is a small or a large industrial complex, research laboratory, or other businesses involving shift works.

Previous methods for creating rotational workforce schedules included interactions between the schedule maker and the algorithm, including defining the length of sequences of consecutive days of working shifts. 

In this method, an algorithm takes into account inputs (or constraints) from the schedule maker and then presents the possible solutions (incl. that all shifts must be filled, working hours per week, minimal resting time, etc.) in a first phase. The schedule maker can then select which solutions are most feasible to proceed with in the second phase, where the final schedules are then constructed and exported.
\end{abstract}

\section{Introduction}
Creating shift work schedules has always been a challenging task, especially such that are equal for all workers and at the same time distributes the shifts evenly and properly to prevent staff burnout. In order to achieve schedules for the workers that treats everyone equally, we will focus on so-called rotational workforce schedules (RWS:s). Rotational workforce schedules means that the schedule rotates after time, and hence, the other option would be static shift schedules.

In this project, we will do some literature review of what has been done in the past to establish methods for creating RWS:s followed by a different method to create them as well as follow the examples of creating them.

\section{Literature review}
In \cite{EfficientRWS}, the framework proposed consists of 4 steps (as quoted from \cite{EfficientRWS} p. 88):
\begin{itemize}
  \item choosing a set of lengths of work blocks (a work block is a sequence of consecutive days of working shifts),
  \item choosing a particular sequence of work and blocks of days-off amongst these that have optimal weekend characteristics,
  \item enumerating possible shift sequences for the chosen work blocks subject to shift change constraints and bounds on sequences of shifts, and
  \item assignment of sequences of shifts to blocks of work while fulfilling the staffing requirements.
\end{itemize}
The framework can thus, as stated in the article, be considered as a semi-automatic way of generating schedules as the algorithm is focused on interactions with the decision-maker who has to choose a fixed set of allowable lengths for work blocks.

In \cite{DecHeuRWS}, an algorithmic framework using a decomposition heuristic was constructed in order to quickly obtain feasible solutions for the rotational workforce scheduling problem. Here, similar to \cite{EfficientRWS}, the decision-maker has to set up a fixed set of work blocks. Their method is very powerful and can construct schedules for a large number of employees in a short amount of time.

In this project, we want to make the amount of interaction required less by allowing an algorithm present all possible solutions and hence giving the schedule maker all possible options for a given number of weeks to cycle over (which in this project is the same as the number of workers).

\section{Computational Approach and Results}
Looking back at what was done in previous work \cite{EfficientRWS}\cite{DecHeuRWS}, we will use an approach similar to what was suggested in \cite{EfficientRWS}: Each person has the same schedule, shifted by one week. Thus, the weeks are continuously shifted by one week for each worker until all workers have had 'each week'. This type of schedule can be created manually by a person by e.g. defining a set of work blocks (\cite{EfficientRWS}\cite{DecHeuRWS}), but in this project an algorithm has been constructed that does the work of finding possible work blocks in order to reduce the workload for generating them. In this approach, we define the term 'shift arrays' which each represent a possible schedule, similar to work blocks. It has been divided into two phases, \textit{Boolean Shift Arrays} (in which boolean shift arrays are generated) and \textit{From Boolean Shift Arrays to a RWS} (in which a selected boolean shift array is shaped into its final RWS layout).

\subsection{Boolean Shift Arrays (phase 1)}
For simplicity, we begin by using boolean shift arrays with 1 meaning that the person works and 0 meaning that the person does not work. In order to impose some constraints on the shift arrays, we define the number of working days per week as $n_{wd}$ and number of weeks to cycle over as $n_W$. This results in a shift array of length $7n_W$. For constructing the different combinations of the boolean array, we use Algorithm~\ref{algorithm1}.

\begin{algorithm}[H]
\SetAlgoLined
shiftarrays = []\;
r = range(self.noofweeks*self.workingdays)\;
pool = tuple(iterable)\;
n = len(pool)\;
\eIf{r $>$ n}{
    loop = False\;
}{
    loop = True\;
}
\While{loop is True}{
    \For{i in reversed(range(r))}{
    \If{indices[i] != i + n - r}{
            break
        }
    }
    \Else{
    loop = False
    }
    \If{loop is True}{
        indices[i] += 1\;
        \For{j in range(i+1, r)}{
            indices[j] = indices[j-1] + 1
        }
        shiftarray = ["0"] * (numberofweeks*workingdays)\;
        \For{ind in tuple(pool[i] for i in indices)}{
            shiftarray[ind] = "1"
        }
        \If{shiftarray is ok}{
        shiftarrays.append(shiftarray)
        }
    }
}
\caption{Generation of Boolean Arrays}
\label{algorithm1}
\end{algorithm}

Note that the "\textbf{if} \textit{shiftarray is ok}" in Algorithm~\ref{algorithm1} is a combination of all constraints discussed in this section and are checked before a shiftarray is allowed to be appended to the list of shiftarrays, hence ensuring that all shiftarrays in the list follow these constraints.

Since each week also resembles a worker, the shift array can be set up as a matrix with 7 columns, each representing the days of a week. The columns can then be summed to achieve the shift occupancy (or how many people are working each shift). Thus, the algorithm only allows shift arrays to pass for which all shifts are occupied by at least one worker, with a shift represented by the first $n_{wd}$ days for each week. In order to extend to not only use single shifts but also 2- or 3-shifts, a simple logical reasoning was added into the algorithm. For $N$ shifts per day, each day has to be filled with at least $N$ workers.

The next input that the algorithm needs is the shift lengths and the weekly working hours per worker, defined as $t_s$ and $t_W$, respectively. However, in order to generate "good schedules", an additional constraint will be needed in order to avoid all working days being clustered together.

\begin{figure}[h]
  \centering\includegraphics[width=0.9\columnwidth]{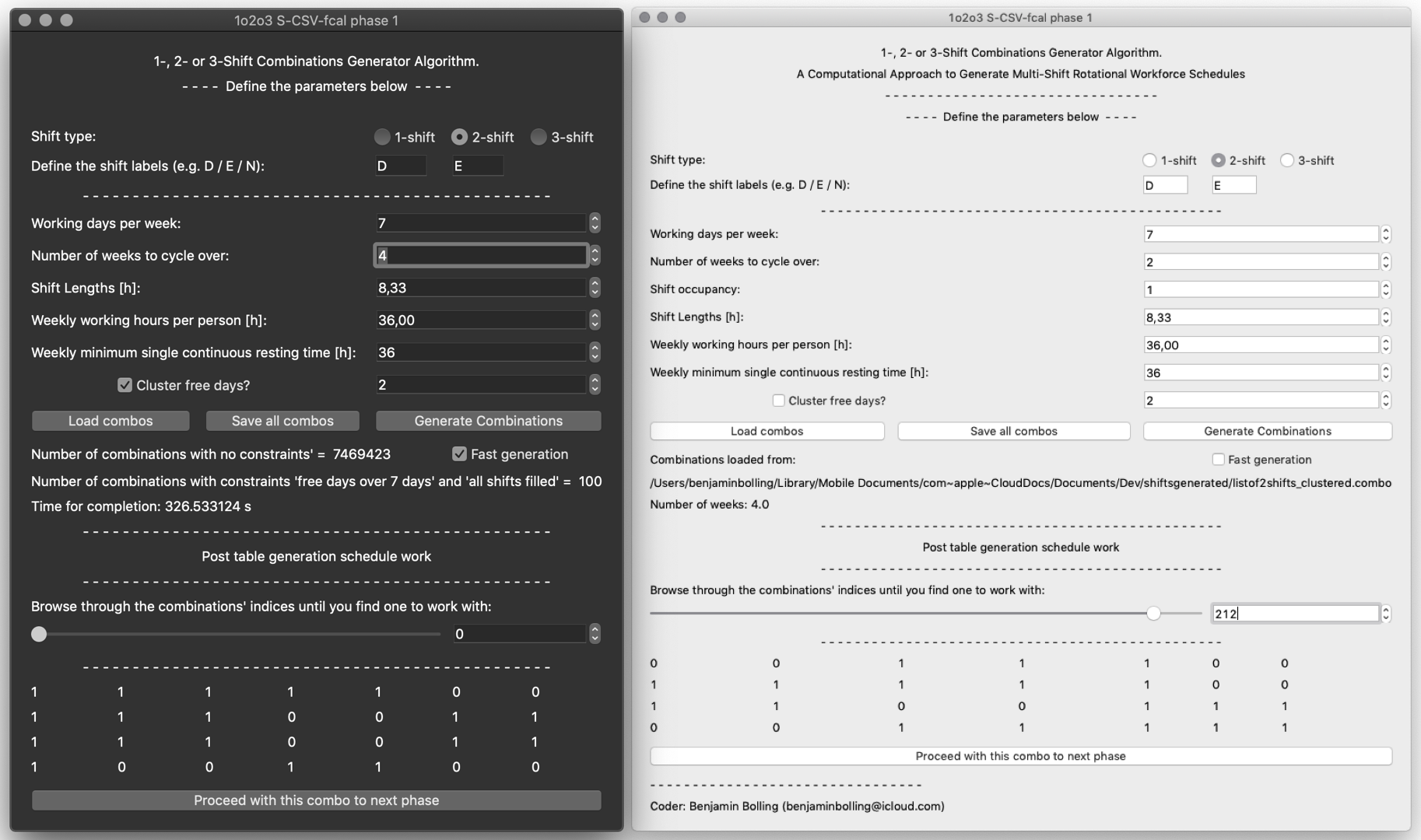}
\caption{The RWS:ing Application's algorithm's "phase 1 GUI" (dark and light themes, left and right, respectively). In the left figure, the combinations have been generated. In the right figure, the combinations have been loaded from a file.}
\label{fig:figure1}
\end{figure}

Hence, the constraint for weekly minimum single continuous resting time is added, defined as $t_r$. The algorithm ensures that all passed shift arrays have at least this many hours of free-time for each week. The number of shifts $n_S$ per shift array is calculated by 
\[
n_S = \mathrm{ceil}\left(\frac{t_W}{t_s}\right).
\]
The reason for using ceiling function and not the floor function is simply the argument that it is better with a couple of more hours than fewer. In order to cluster days off ($n_cf$), the algorithm's GUI has an optional additional constraint that serves this purpose and simply does not allow shift arrays with 0:s in clusters less than this through.

By using the input $n_Wn_{wd}$ as an iterable and $n_S$ as the length of subsequences of elements from the iterable, we use the same methodology as \textit{itertools} \cite{Itertools} module in Python to create each shift array. The other inputs are used as constraints on whether the shift array should be appended to the array of shift arrays or trashed. The reasoning for not using the built-in module itertools.combinations \cite{Itertools} is that it returns all array combinations it could find. Without the constraints, the returned arrays become too large for a normal up-to-date computer's internal memory to handle.

With this, the final result is an array of shift arrays in which each shift array is filled with $7n_S$ 1:s and $n_W(7-n_S)$ 0:s whilst obeying the above mentioned constraints. The number of possible combinations $C$ using Algorithm~\ref{algorithm1} can be expressed as:
\begin{equation}
    C = \frac{(n_Wn_{wd})!}{n_S! (n_Wn_{wd}-n_S)!}.
\end{equation}

\begin{table}[h]
\centering
\caption{Parameters selected for the generation of a N-shift RWS.}
\label{tab:table1}
\begin{tabular}{| cccccccc |}
\hline\noalign{\smallskip}
$N$ & $n_{wd}$ & $n_W$ & $t_S$ & $t_W$ & $t_r$ & $n_cf$ & Shift types' labels\\
\noalign{\smallskip}
\hline\noalign{\smallskip}
2 & 7 & 4 & 8.33 & 36.00 & 36 & 2 & D, E\\
\noalign{\smallskip}
\hline
\end{tabular}
\end{table}

The parameters selected for the RWS is defined in Table~\ref{tab:table1}. These values are then reflected in the algorithm's GUI for phase 1 can be seen in Figure~\ref{fig:figure1} to the left. Note that the generated shift arrays can be browsed through using the slider or the numerical input field and that the shift arrays are constructed such that each week or worker (depending on the viewing angle) is represented in separate rows.

\subsection{From Boolean Shift Arrays to RWS (phase 2)}
\begin{figure}[h]
  \centering\includegraphics[width=0.98\columnwidth]{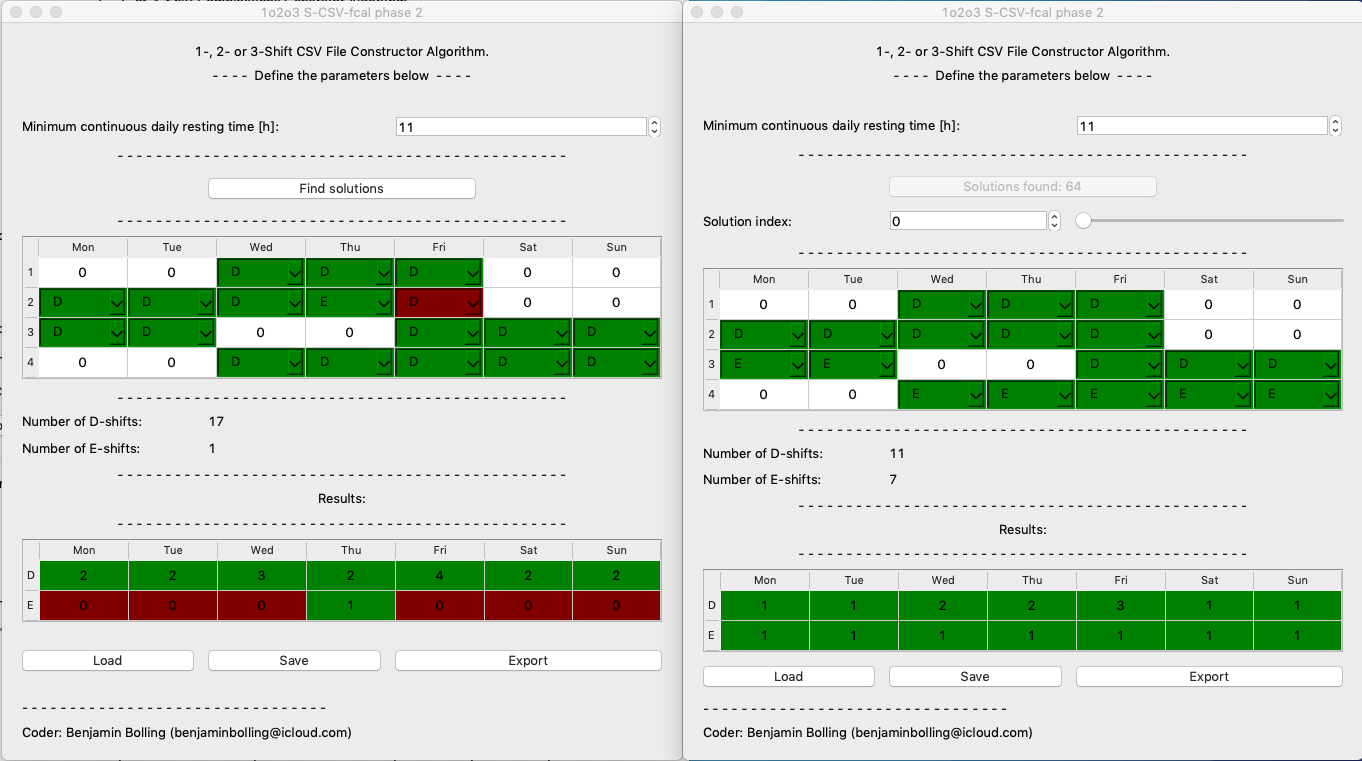}
\caption{The RWS:ing Application's algorithm's "phase 2 GUI" as launched from the "phase 1 GUI" and with the second Thursday's shift changed to an evening shift (left) and after finding solutions, showing the first solution (right).}
\label{fig:figure2}
\end{figure}
In this phase, we have generated a new list of combinations with free days clustered in pairs, and then we chose to proceed with combination \#212 since it has two out of four weekends off (note the zeroes in the bottom table in Figure~\ref{fig:figure1} to the right). By proceeding, the "phase 2 GUI" is launched with the selected array as input as can be seen in Figure~\ref{fig:figure2} to the left.

The free days are all represented by zeroes whilst all other shifts (ones) are converted to the first defined shift type label. For $N>1$, each shift can be replaced by another shift via dropdown menus. The GUI shows the number of shifts of each type each week or worker has and a table with the results, i.e. number of worker per shift and day. Shifts that are occupied have green background whilst shifts that are unoccupied have a red background.

If the continuous resting time between two assigned shifts is too low, the background colour of the second shift becomes red (e.g. a Friday day-shift after a Thursday evening-shift if the continuous resting time has to be at least 11 hours, as shown in Figure~\ref{fig:figure2} to the left). With enough resting time in between shifts, the background of the second shift is be green. 

Pressing the \textit{Find solutions} results in what is shown in Figure~\ref{fig:figure2} (right figure). A schedule can also be constructed completely by hand, but note that the algorithm will find all possible combinations. The algorithm is a simple Cartesian Product calculator, in which each set is a list of shifts (1 = Day, 2 = Evening, etc.) with one set per working day:
\[
\mathrm{combinations} =
\underbrace{
\begin{pmatrix} 1 \\ 2 \\ \vdots\end{pmatrix}
\times
\begin{pmatrix} 1 \\ 2 \\ \vdots\end{pmatrix}
\times ... \times 
\begin{pmatrix} 1 \\ 2 \\ \vdots\end{pmatrix}
}
_{n_{wd}} = 
\begin{cases}
  [1\;1\;\cdot\;\cdot\;\cdot\;1]\\
  [1\;1\;\cdot\;\cdot\;\cdot\;2]\\
  \;\;\;\;\;\;\;\vdots\\
  [2\;2\;\cdot\;\cdot\;\cdot\;1]\\
  [2\;2\;\underbrace{\cdot\;\cdot\;\cdot}_{n_{wd}}\;2]
\end{cases}
\]
where each array in the resulting product is considered as a possible shift schedule matrix. Imposing constraints (resting time between shifts and ensuring all shifts are filled) on each combinations results in solutions from which the user can choose between.

Since all combinations are stored in a matrix form before different combinations are removed from the final solutions matrix, large datasets require severe amount of internal memory for the Cartesian Product method to work. For this, a controlling script has been implemented which calculates a pre-estimate of required internal memory. The required internal memory for different operations can be roughly calculated by
\[
\mathrm{IM} \approx N_{C}n_{wd} = N^{n_{wd}}n_{wd},
\]
returning the memory demand $\mathrm{IM}$ in bytes and where $N_C = N^{n_{wd}}$ is the total number of combinations (without any constraints imposed).

If the estimated expected internal memory requirement for an operation exceeds 1Gb, the user is prompted whether to continue with the default Cartesian Product method or to use a less internal memory demanding recursive method. The recursive method can be simplified as shown in Algorithm~\ref{algorithm2}.

\begin{algorithm}[H]
\SetAlgoLined
def recursiveCaPr(N,shifts,arrays,array,level):\;
    \For{m in range(1,len(shifts))}{\;
        \For{n in range(level-1,N)}{\;
            \If{array[n] != shifts[m]}{\;
                array2 = array.copy()\;
                array2[n] = shifts[m]\;
                matrixOut = insertFreeDaysInSolutionMatrix(array2)\;
                \If{matrixOut not in arrays}{\;
                    \If{checkifshiftsOK(matrixOut) is True}{\;
                        arrays.append(matrixOut)\;
                    }
                    \If{level < N:}{\;
                        arrays = recursiveCaPr(N,shifts,arrays,array2,level+1)\;
                    }
                }
            }
        }
    }
    return arrays\;
arrays = [[shifts[0]] * N]\;
solutions = recursiveCaPr(N,shifts,arrays,arrays[0],level)\;
\caption{Finding solutions to a Cartesian Product of length N using a recursive function.}
\label{algorithm2}
\end{algorithm}

\section{Benchmarking results}
\subsection{Benchmarking Computer Specifications}
The algorithm benchmarking was done on an Apple MacBook Pro with the specifications as defined in Table~\ref{tab:computerSpecs}.
\begin{table}[h]
\centering
\caption{Benchmarking computer specifications.}
\label{tab:computerSpecs}
\begin{tabular}{| r l |}
\hline\noalign{\smallskip}
Computer type: & Apple MacBook Pro (13-inch, 2019)\\
OS: & macOS Mojave v. 10.14.6\\
Processor: &  2.8 GHz Intel Core i7 processor\\
Internal Memory: & 16 GB 2133 MHz LPDDR3\\
Graphics Card: &  Intel Iris Plus Graphics 655 1536 MB\\
\noalign{\smallskip}
\hline
\end{tabular}
\end{table}

\subsection{Phase 1 - Construction of  Combinations as Boolean Arrays}
In the GUI, there is a "fast generation" checkbox which stops the algorithm from further calculations once the first 100 solutions have been found. This way, computation time can be lowered (in comparison to "full generation" which will go through all possible solutions from the boolean array). For our example, the time it took to complete decreased from 508.7 s (for a full generation) to 24.55 s (for the full generation) (see Table~\ref{tab:fullBenchmarking}), which is a decrease in time by 95\%. 

We use the parameters defined in Table~\ref{tab:table1}, with the exception of $N$ and Shift types' labels. Note that for Table~\ref{tab:fullBenchmarking} and Table~\ref{tab:fullCombos}, the number (\#) of weeks given is the minimum amount of weeks required for a full shift cycle in order to find solutions for the N-shift problems (with $N = 1,2,3$ for single-, two- and three-shifts, respectively). The free days clustering option is not selected for the benchmarking. 

\begin{table}[h]
\centering
\caption{Benchmarking for fast and full generation of the Boolean Arrays (as defined in Section 3.1 for Phase 1).}
\label{tab:fullBenchmarking}
\begin{tabular}{| r | c | r r |}
\hline\noalign{\smallskip}
Type: & \# of weeks: & Time (fast) [s]: & Time (full) [s]:\\
\noalign{\smallskip}
\hline\noalign{\smallskip}
Single-shift, 5 days/week   & 1 & 7.224e-05 & 7.224e-05\\
Single-shift, 7 days/week   & 2 & 1.497e-02 & 5.211e-02\\
Two-shift, 7 days/week      & 4 & 24.55 & 508.7\\
Three-shift, 7 days/week    & 5 & 3 087 & 6.627e+04\\
\noalign{\smallskip}
\hline
\end{tabular}
\end{table}

\begin{table}[h]
\centering
\caption{Number of combinations and solutions found for full generations of the Boolean Arrays (as defined in Section 3.1 for Phase 1).}
\label{tab:fullCombos}
\begin{tabular}{| r | c | r r |}
\hline\noalign{\smallskip}
Type: & \# of weeks: & Combinations: & Solutions:\\
\noalign{\smallskip}
\hline\noalign{\smallskip}
Single-shift, 5 days/week   & 1 & 1 & 1\\
Single-shift, 7 days/week   & 2 & 2 002 & 462\\
Two-shift, 7 days/week      & 4 & 13 123 110 & 1 668 226\\
Three-shift, 7 days/week    & 5 & 1 476 337 800 & 11 383 225\\
\noalign{\smallskip}
\hline
\end{tabular}
\end{table}

\begin{figure}[h]
  \centering\includegraphics[width=0.94\columnwidth]{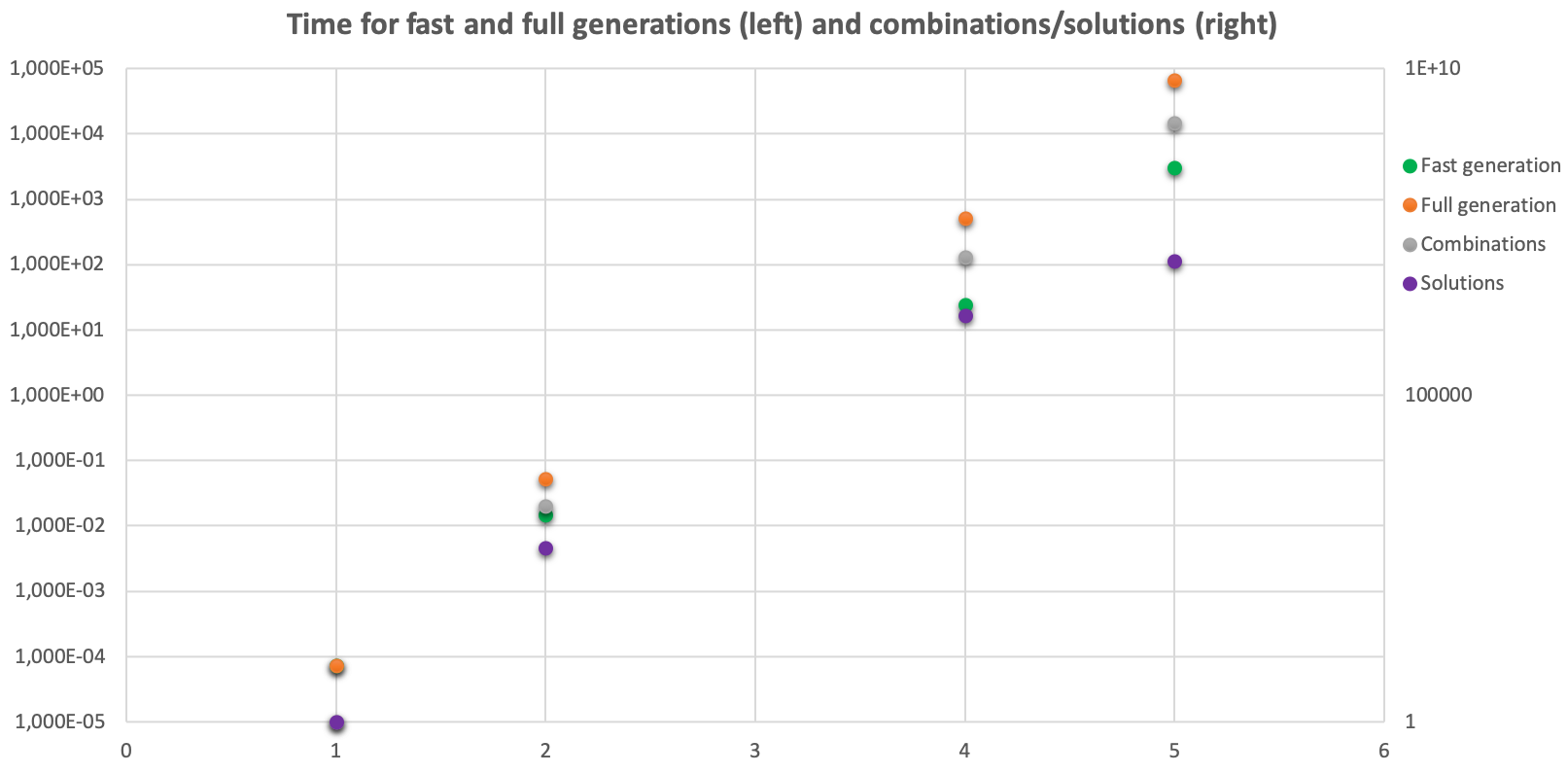}
\caption{The benchmarking results in respect of time for fast- and full generation of the boolean arrays (on the left vertical axis), and the number of combinations gone through and the solutions found (on the right vertical axis).}
\label{fig:figure4}
\end{figure}

Plotting the benchmarking results, we find the logarithmic graph given in Figure~\ref{fig:figure4}. As can be seen, the computation time $T_C$ increases exponentially with the number of weeks in a shift cycle on average in accordance with
\[
T_C(full) = \exp(5.046\times n_W)\times9\times10^{-7}
\]
and
\[
T_C(fast) = \exp(4.254\times n_W)\times2\times10^{-6}
\]
for the full and fast generations on, respectively.

\subsection{Phase 2 - Finding Solutions for a given Combination}
If the given combination has only a single shift specie, there is one solution for the given combination. If there are more than one shift specie, multiple solutions may be found. The main impact on time consumption is the number of combinations $N_C$. Limiting factors are not limited to time only but also on the internal memory due to that a Cartesian Product method is used, meaning all combinations are stored as list objects. Some values have been timed and calculated in Table~\ref{tab:phase2} using the Cartesian Product method.

\begin{table}[h]
\centering
\caption{Benchmarking for Phase 2: Time and estimated internal memory (IM) required for obtaining all combinations and solutions for different $n_{wd}$ and $N$ using the Cartesian Product method.}
\label{tab:phase2}
\begin{tabular}{| r | c c | r r | r r |}
\hline\noalign{\smallskip}
Type ($N$): & $n_{wd}$: & $n_{W}$: & Combinations: & Solutions: & IM: & Time [s]:\\
\noalign{\smallskip}
\hline\noalign{\smallskip}
2-shift & 14    & 3 & 16 384 & 7 & 229.38 kB & 0.2963\\
2-shift & 18    & 4 & 262 144 & 64 & 4.7186 MB & 5.843\\
3-shift & 14    & 3 & 4 782 969 & 0 & 66.96 MB & 92.54\\
3-shift & 18    & 4 & 387 420 489 & - & 6.9736 GB & -\\
\noalign{\smallskip}
\hline
\end{tabular}
\end{table}

\section{Example Procedure}
\begin{figure}[h]
  \centering\includegraphics[width=\columnwidth]{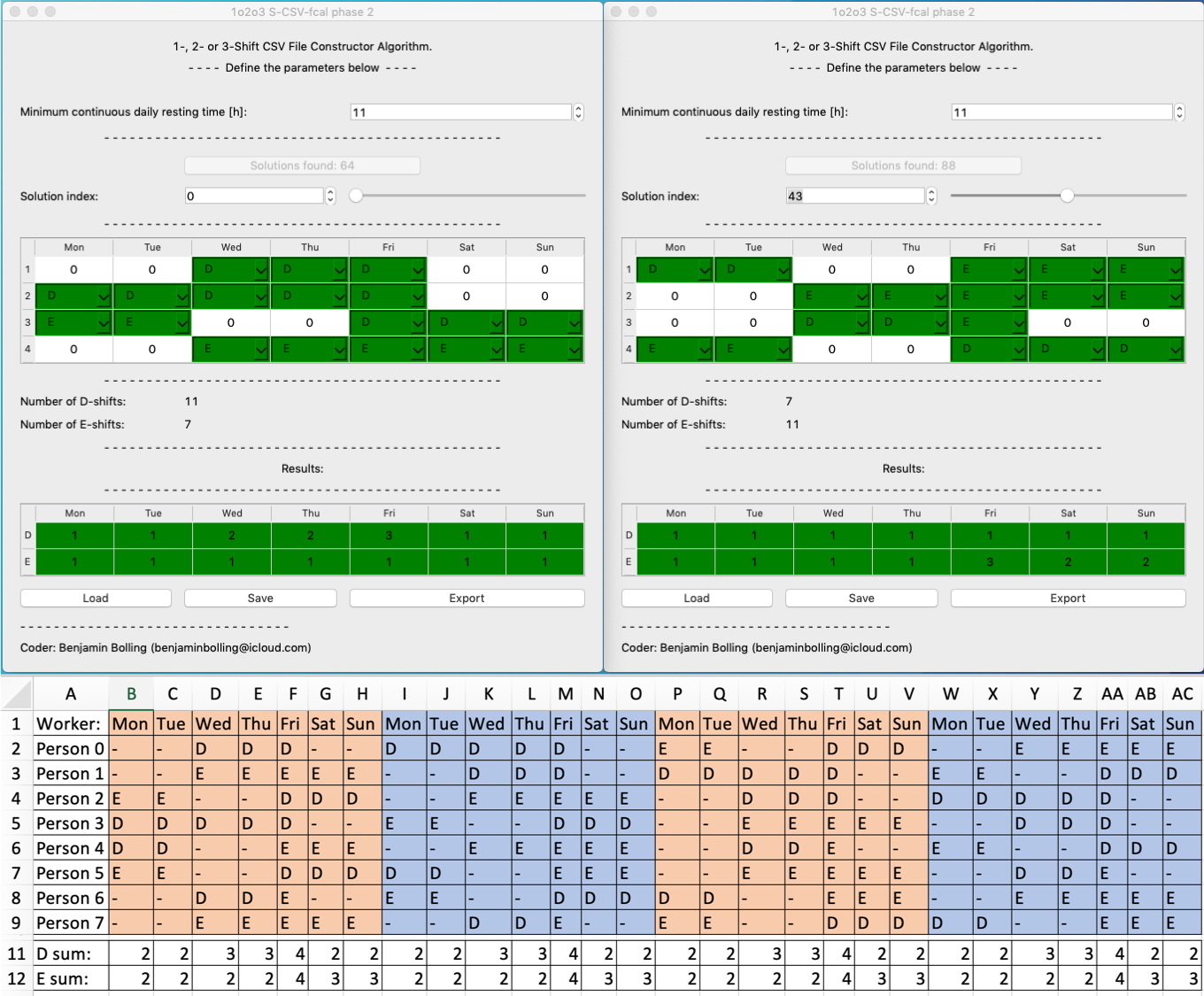}
\caption{Example procedure results. Phase 1 was performed using the parameters defined in Table~\ref{tab:table1} and with free days clustered in pairs. Two boolean combinations were selected, followed by finding and selecting a solution for both, and then exported as CSV and imported in Microsoft~Excel~(2018).}
\label{fig:figure3}
\end{figure}

We again use the parameters defined in Table~\ref{tab:table1} and then select to cluster the free days in pairs. The boolean combinations selected in phase 1 for phase 2 are \#212 and \#43. In phase 2, the solutions selected are \#0 and \#43 for the boolean combinations \#212 and \#43, respectively. These solutions were then exported as .CSV files, imported into a new spreadsheet in Microsoft~Excel~(2018) followed by adding colours to highlight and differentiate each week as well as lines for all rows and columns. The sum of each shift specie per day has also been calculated in the spreadsheet (using the COUNTIF(cells;shift) formula of Microsoft~Excel~(2018)), see Figure~\ref{fig:figure3}.

Thus it can be concluded that in this example procedure two different 4-week (or 4-people) rotational workforce schedules were generated and then combined together. In the first schedule (persons 0-3, Figure~\ref{fig:figure3}), each staff has two "normal office-hours" weeks with the weekends off, followed by two weeks with weekends working. The second schedule (persons 4-7, Figure~\ref{fig:figure3}) has more evening and weekend shifts, which can be considered as more suited for e.g. students or people working some extra hours. These two schedules have then been combined in order to obtain an a rotational workforce schedule filled as evenly as possible with the exception of Fridays. Fridays have been selected such that all shift workers are scheduled to work, suitable for e.g. activities when \textit{all hands on deck} is required. This schedule has also ensured that each shift is occupied by a minimum of two shift workers.

\section{Conclusions}
In this article, we have demonstrated that the constructed algorithm can generate schedules for different number of weeks to cycle over. The current issue is that the computational complexity (and hence the computation time) increases with the number of weeks, as can be seen in Table~\ref{tab:fullCombos} and Figure~\ref{fig:figure4}. This means that for a higher amount of weeks in a shift cycle, this application will need development in order to have more efficient ways of finding the solutions and/or deployment of the application onto super-computers for generating the Boolean Arrays.

For up to 5 weeks in a shift cycle it is possible to use a general-purpose computer such as the benchmarking Apple MacBook Pro with specifications defined in Table~\ref{tab:computerSpecs}. 
It has thus been demonstrated that the application can be used to generate 1, 2 and 3-shift schedules. Future development plans include adding an automated assignment function of shift types in phase 2, which would further strengthen the usability of this application.

\section{Conflict of interest statement}
On behalf of the author, the corresponding author states that there is no conflict of interest.

\end{document}